\documentclass{aa}

\usepackage{txfonts}
\usepackage{graphicx}
\usepackage{rotating}

\begin{document}

\title{ Scattered H$\alpha$ emission from a large
        translucent cloud G294-24}
\author{ K.\ Lehtinen \and M.\ Juvela \and K.\ Mattila }

\institute{Observatory, T\"ahtitorninm\"aki, P.O.
           Box 14, 00014, University of Helsinki, Finland
           \thanks{\emph{Present address:} Department of Physics,
           Division of Geophysics and Astronomy,
           P.O.\ Box 64,
           FI-00014 University of Helsinki, Finland } }

\date{Received / accepted}

\offprints{K.\ Lehtinen (kimmo.lehtinen@helsinki.fi)}

\titlerunning{}
\authorrunning{K.\ Lehtinen et~al.} 
\abstract
{}
{ We study an undocumented large translucent cloud, detected by means
of its enhanced radiation on the SHASSA (Southern H-Alpha Sky Survey
Atlas) survey. We consider whether its excess surface brightness can
be explained by light scattered off the dust grains in the cloud, or
whether emission from in situ ionized gas is required. In addition, we
aim to determine the temperature of dust, the mass of the cloud, and
its possible star formation activity. }
{ We compare the observed H$\alpha$ surface brightness of the cloud
with predictions of a radiative transfer model. We use the WHAM
(Wisconsin H-Alpha Mapper) survey as a source for the Galactic
H$\alpha$ interstellar radiation field illuminating the cloud.  Visual
extinction through the cloud is derived using 2MASS $J$, $H$, and $K$
band photometry. We use far-IR ISOSS (ISO Serendipitous Survey), IRAS,
and DIRBE data to study the thermal emission of dust. The LAB (The
Leiden/Argentine/Bonn Galactic HI Survey) is used to study 21\,cm HI
emission associated with the cloud. }
{ Radiative transfer calculations of the Galactic diffuse H$\alpha$
radiation indicate that the surface brightness of the cloud can be
explained solely by radiation scattered off dust particles in the
cloud.  The maximum visual extinction through the cloud is about
1.2\,mag. The cloud is found to be associated with 21\,cm HI emission
at a velocity $\sim-9$\,km\,s$^{-1}$.  The total mass of the cloud is
about 550--1000\,M$_{\sun}$. There is no sign of star formation in
this cloud. The distance of the cloud is estimated from the Hipparcos
data to be $\sim 100$\,pc. }
{}
\maketitle

\section{Introduction}
del Burgo \& Cambr\'esy (\cite{burgo06}) were the first to report on
detection of diffuse H$\alpha$ emission in a molecular cloud.  They
detected an excess surface brightness over the cloud \object{LDN1780} of
intensity $\sim$1--4 rayleigh (one rayleigh (R) being equivalent to
$2.4\,10^{-7}$\,ergs\,cm$^{-2}$\,s$^{-1}$\,sr$^{-1}$ at the wavelength
of H$\alpha$ emission).  They interpreted the surface brightness as a
result of enhanced in situ cosmic ray ionization.  However, Mattila
et~al.\ (\cite{mattila07}) showed that the H$\alpha$ surface
brightness observed in LDN1780 can be explained solely in terms of
scattered H$\alpha$ radiation.  In addition, Mattila et~al.\ found
several other molecular clouds that had excess H$\alpha$ emission
relative to the surroundings of the cloud. In some cases, a cloud was
not detected in H$\alpha$ images, although other physically similar
clouds do.  These observations can be naturally explained in the
framework of scattered H$\alpha$ radiation by varying the proportions
of general diffuse in situ H$\alpha$ emission either in front of or
behind the dust cloud.

While comparing the all-sky H$\alpha$ and 100\,$\mu$m IRAS maps, we
noticed a large cloud visible in both maps. The Galactic coordinates
of the cloud are $l \approx 294\degr$, $b \approx -24\degr$.  The
maximum excess surface brightness of H$\alpha$ is about 2.4\,R.  With
a size of about 1.4\degr$\times$4.9\degr, the cloud is the largest
dust cloud visible in the light of scattered H$\alpha$ emission known
to us. Hereafter the cloud is called G294-24.

\section{Observations and calculations}

\begin{figure*}
\resizebox{\hsize}{!}{\includegraphics{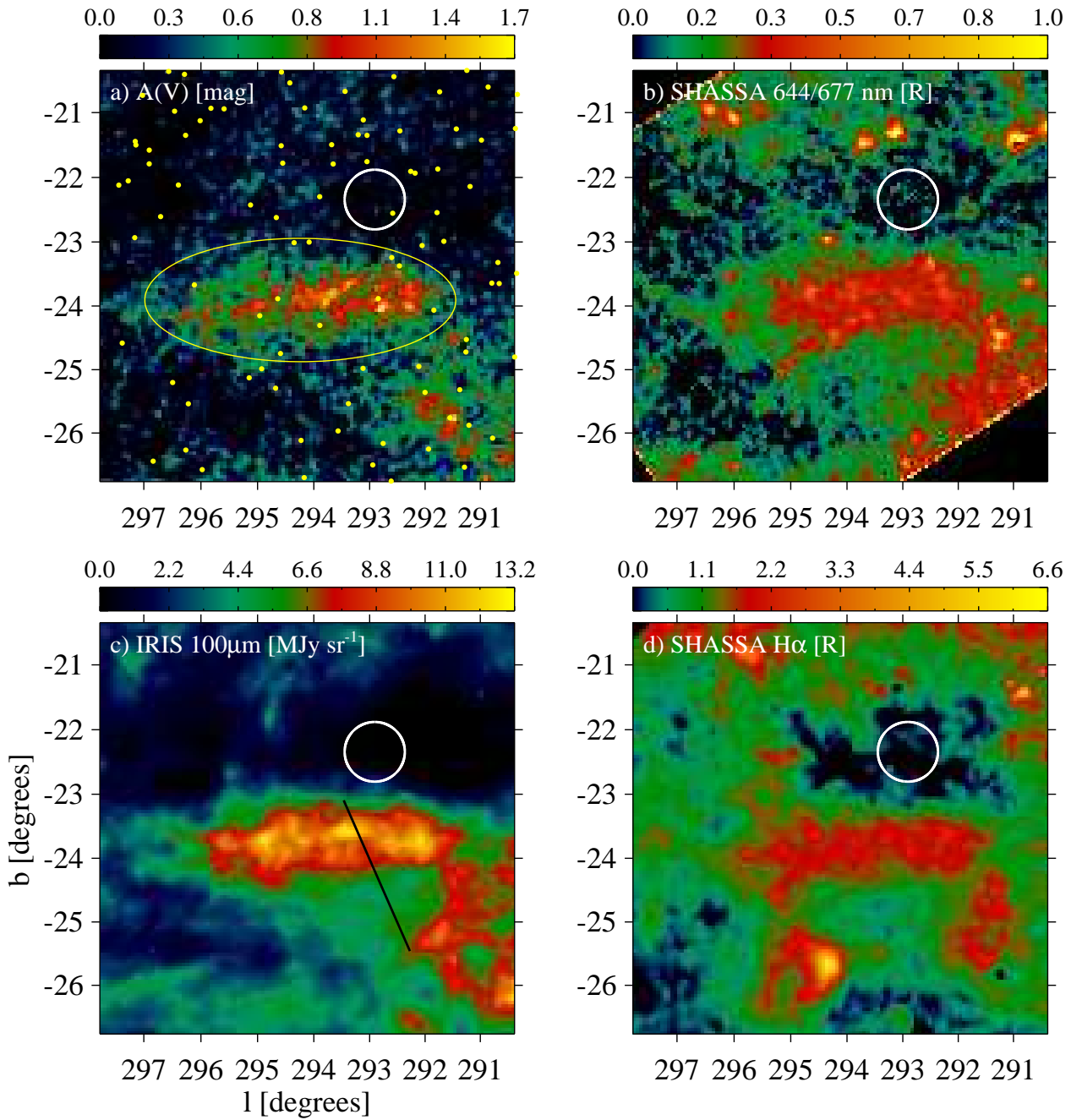}}
\caption{ Map of visual extinction ({\bf a}), map of 644/677\,nm
          continuum intensity ({\bf b}), map of IRAS 100\,$\mu$m
          intensity ({\bf c}), and map of H$\alpha$ intensity ({\bf
          d}).  The value of background at each wavelength was
          determined within the circle, and then subtracted from the
          maps.  The line in panel {\bf c} shows the location of the
          slew of the ISO Serendipitous Survey (ISOSS).  In panel {\bf
          a}, dots indicate the Hipparcos stars, while the ellipse
          delineates the extent of the cloud. }
\label{fig1}
\end{figure*}

\begin{figure}
\resizebox{\hsize}{!}{\includegraphics{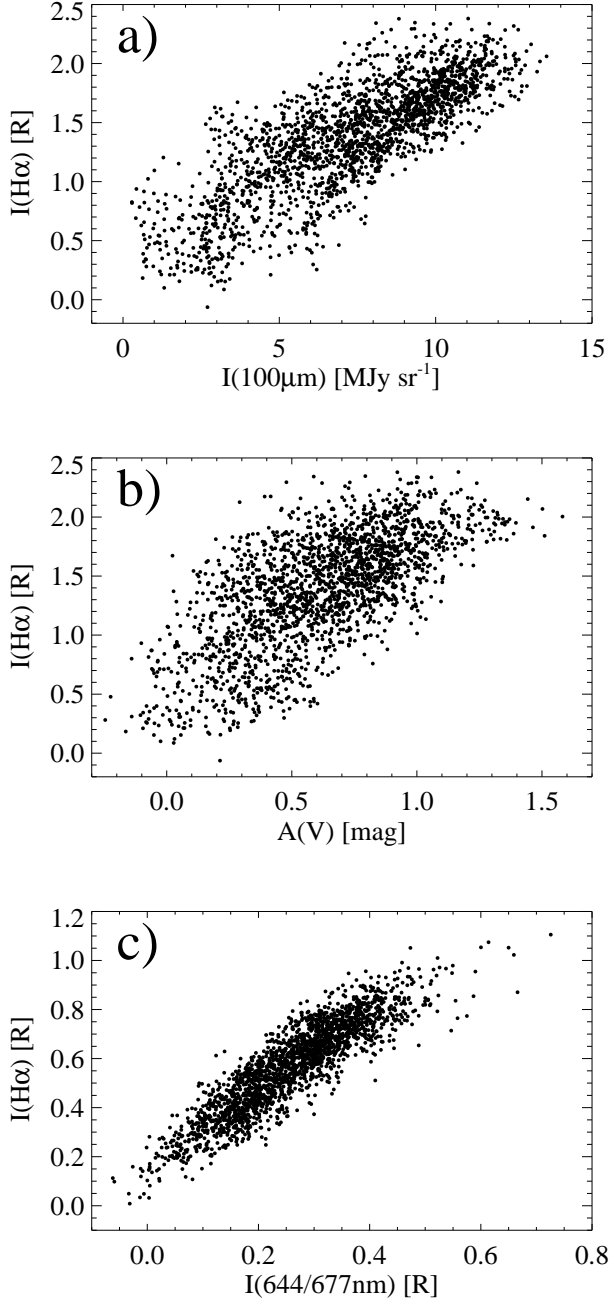}}
\caption{ The observed H$\alpha$ surface brightness as a function of
  IRAS 100\,$\mu$m intensity ({\bf a}), visual extinction ({\bf b})
  and 644/677\,nm intensity ({\bf c}). The background values have been
  subtracted from each dataset. Note that the H$\alpha$ data in panels
  {\bf a} and {\bf b} are from Finkbeiner (\cite{finkbeiner03}), while
  in panel {\bf c} the data are from Gaustad et~al.\
  (\cite{gaustad01}). }
\label{fig2}
\end{figure}

\begin{figure}
\resizebox{\hsize}{!}{\includegraphics{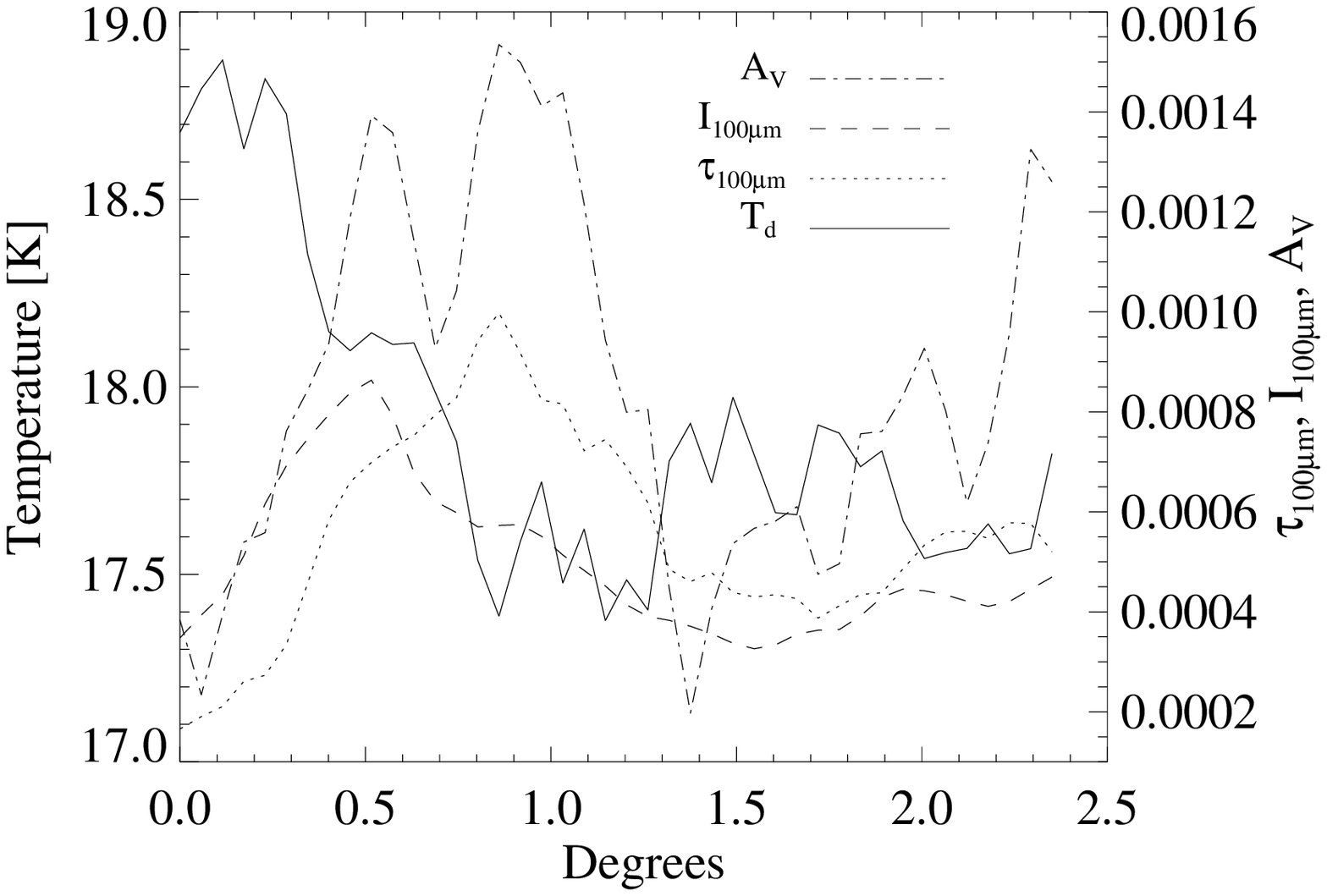}}
\caption{ Solid line: temperature of dust, $T_d$, along the ISOSS slew
  shown in Fig.~\ref{fig1}c, derived from 100\,$\mu$m IRAS and
  170\,$\mu$m ISOSS data.  Dotted line: 100\,$\mu$m optical depth,
  $\tau(100\mu$m), along the same slew. Dashed line: 100\,$\mu$m
  surface brightness, $I(100\mu$m), along the same slew, divided by
  1.4$\times$10$^4$. Dash-dotted: visual extinction, $A_V$, along the
  same slew, divided by 7$\times$10$^2$. }
\label{fig3}
\end{figure}

Dust grains in the cloud G294-24 are seen in the light of scattered
and emitted radiation. In addition, dust grains cause extinction of
light of those stars which are located behind the cloud. 
Atomic hydrogen gas in the cloud is expected to be seen in the 21\,cm 
spin-flip line. 
We are studying these components by utilizing different data archives. 

\subsection{H$\alpha$ surface brightness}
We obtained the H$\alpha$ data from the SHASSA (Southern H-Alpha Sky
Survey Atlas) survey (Gaustad et~al.\ \cite{gaustad01}), which is
incorporated into the all-sky composite map by Finkbeiner
(\cite{finkbeiner03})\footnote{http://skymaps.info}. The data
were re-gridded into a regular grid in Galactic coordinate system,
using a pixel size of 3.4\arcmin$\times$3.4\arcmin, which gives the
same pixel area as the original HEALPix data.  The H$\alpha$ data have
a spatial resolution of FWHM=6\arcmin. The intensities are given in
units of rayleigh (R).

\subsection{644/677\,nm surface brightness}
The SHASSA survey includes continuum images taken with a dual-band
(644 \& 677\,nm) notch filter. In the original SHASSA survey, these
images were used to subtract continuum emission from the H$\alpha$
images.  In the case of dust clouds, these continuum images detect
diffuse interstellar radiation scattered off the dust clouds.  With a
pixel size of 47.6\arcsec, the low surface brightness of the cloud is
superimposed by numerous undersampled stars, which cannot be removed
by fitting the stars with the point spread function of the
instrument. To remove the stars, we replaced each map pixel with a
mean value over a 9$\times$9 pixel area around the pixel in question,
using only those pixels that occupy the lowest 30\% of the intensity
histogram.  The resolution of the image was then about 7.2\arcmin.
Figure~\ref{fig1}b shows the continuum image after most of the stars
were removed. Some residuals remain in place of the brightest stars.
The intensity of the continuum surface brightness in SHASSA for the
644/677\,nm filter is given in rayleigh units. This has been scaled
for the purpose of background subtraction from the H$\alpha$ filter
band. The physical units for the continuum surface brightness are
R/$\AA$ and the value depends on the width of the filter. Thus, we
cannot compare the absolute intensities of the 644/677\,nm and
H$\alpha$ images from the SHASSA survey.

\subsection{IRAS/IRIS and ISOSS (ISO Serendipity Survey) far-IR data}
We used the 100\,$\mu$m IRIS data and the 170\,$\mu$m ISOSS data to
derive equilibrium temperature and column density of the 'big
classical' dust grains in the cloud.  IRIS data set
(Miville-Desch\^enes \& Lagache \cite{miville05}) is an improved
version of the all-sky IRAS/ISSA data.

The 170\,$\mu$m slew data of the ISOPHOT instrument aboard the ISO
satellite were assembled into the ISOPHOT Serendipity Survey data set
by Stickel et~al.\ (\cite{stickel07}). It covers $\sim 15$\% of the
sky, mapped with a grid size of 22\farcs4, and a few slews cover our
cloud. The one that we used in our data analysis is shown in
Fig.~\ref{fig1}c.

\subsection{COBE/DIRBE far-IR data}
Because of the large size of the cloud, it is resolved in DIRBE data,
which have a beam size of about 0.7\degr.  We thus used DIRBE data at
100\,$\mu$m, 140\,$\mu$m, and 240\,$\mu$m to obtain another estimate
of the temperature and column density of the 'big classical' dust
grains.

\subsection{Visual extinction}
We compiled an extinction map of the cloud by applying the NICER (Near
Infrared Color Excess Revised) method of Lombardi \& Alves
(\cite{lombardi01}), which is an optimized color excess technique
using data in three bands simultaneously. Our data comprise $J$, $H$,
and $K_s$ band magnitudes from the 2MASS archive.  The intrinsic
colors of stars were determined in an area having a local minimum in
the 100\,$\mu$m IRIS map, located at $l=292.0\degr$, $b=-22.3\degr$.

The grid used in the extinction map is the same as that of the
H$\alpha$ map. The extinction value at each grid point is a weighted
mean of the individual extinctions of stars obtained by using a
Gaussian with a width of FWHM=6\arcmin\, as a weighting function.

\subsection{Hipparcos data}
Owing to the large size and low extinction of the cloud, the Hipparcos
data (ESA \cite{esa97}) enable us to estimate the distance of the cloud. 
We collected Hipparcos data over $5\degr\times5\degr$ area towards the cloud.
The intrinsic colors of stars as a function of spectral and luminosity
class were used to derive a color excess $E(B-V)$ for each Hipparcos
star in the area.

\subsection{Hydrogen 21\,cm line emission data}
The ``The Leiden/Argentine/Bonn (LAB) Survey of Galactic HI'' dataset
(Kalberla et~al.\ \cite{kalberla05} ; Bajaja et~al.\ \cite{bajaja05})
was used to search for hydrogen emission related to the cloud.  The
angular resolution of the LAB survey is $HPBW\approx0.6\degr$.

\subsection{Subtraction of background sky}
When studying extinction, scattering and emission associated solely
with the cloud G294-24, we need to subtract the background, determined
in the immediate vicinity of G294-24.  Figure~\ref{fig1} shows the
circular area that was used to estimate the background for each
dataset.  The background values of 100\,$\mu$m intensity, extinction
$A_V$, and H$\alpha$ brightness are about 4.1\,MJy\,sr$^{-1}$,
0.5\,mag, and 1.0\,R, respectively.

\section{Results}

\begin{figure}
\resizebox{\hsize}{!}{\includegraphics{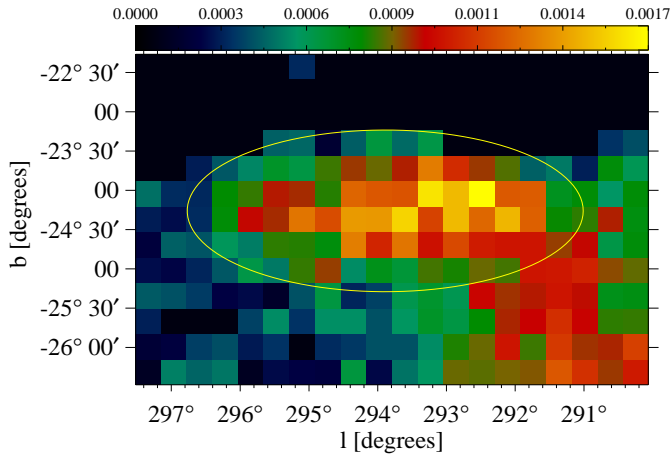}}
\caption{ 100\,$\mu$m optical depth, derived from 100\,$\mu$m,
          140\,$\mu$m, and 240\,$\mu$m DIRBE data. The ellipse, which
          is the same as in Fig.~\ref{fig1}, delineates the area that
          was used to determine the total mass of the cloud. }
\label{fig4}
\end{figure}

\begin{figure}
\resizebox{\hsize}{!}{\includegraphics{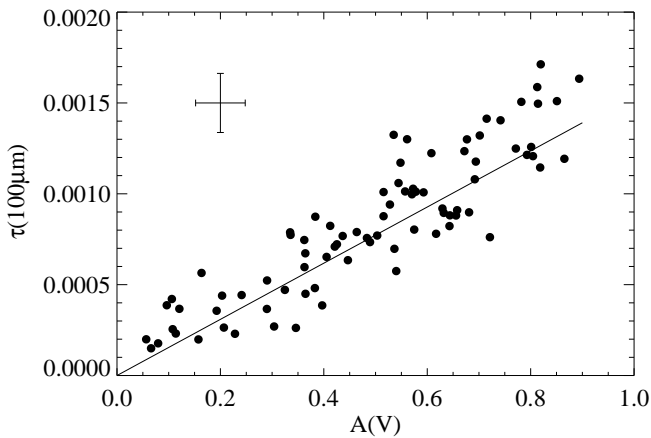}}
\caption{ Relation between 100\,$\mu$m  optical depth and visual
          extinction. Typical 1-$\sigma$ errors are shown. }
\label{fig7}
\end{figure}

\begin{figure}
\resizebox{\hsize}{!}{\includegraphics{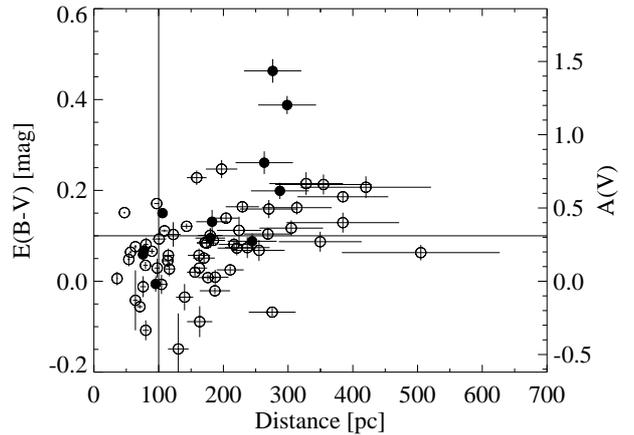}}
\caption{ The reddening $E(B-V)$ of Hipparcos stars as a function of
  their distances for the stars marked in Fig.~1{\bf a}. The stars
  that are shown as filled circles are within the ellipse in
  Fig.1{\bf a}. The vertical line is at a distance of 100\,pc. }
\label{fig5}
\end{figure}

\begin{figure}
\resizebox{\hsize}{!}{\includegraphics{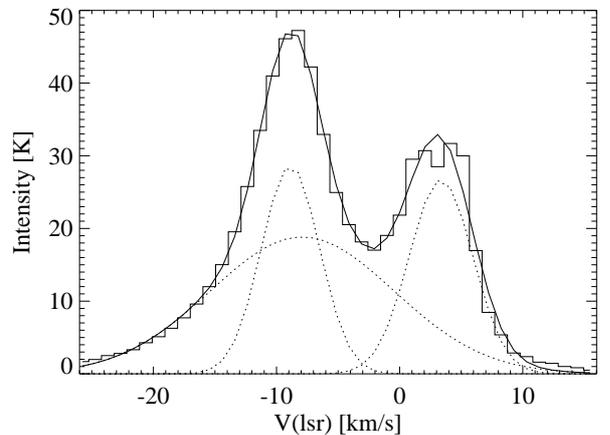}}
\caption{ Spectrum of 21\,cm hydrogen line at the position
          $l=294.0\degr$, $b=-24.5\degr$. Histogram shows the observed
          spectrum, the three Gaussian functions fitted to the
          observed spectrum are shown as dotted lines, while the solid
          line shows the combination of the three Gaussians. The
          brightest component is related to the cloud G294-24. }
\label{fig8}
\end{figure}

\begin{figure}
\resizebox{\hsize}{!}{\includegraphics{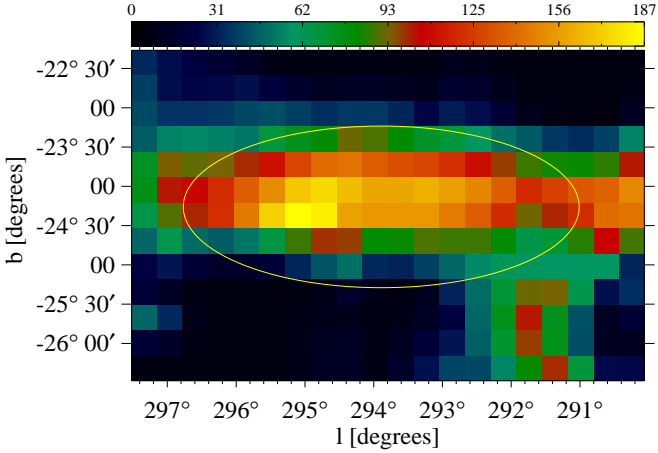}}
\caption{ Map of line area ([K\,km\,s$^{-1}$]) of 21\,cm HI emission
  from the cloud G294-24. The ellipse, which is the same as in
  Fig.~\ref{fig1}, delineates the area used to determine the atomic
  hydrogen mass of the cloud. }
\label{fig9}
\end{figure}

\begin{figure}
\resizebox{\hsize}{!}{\includegraphics{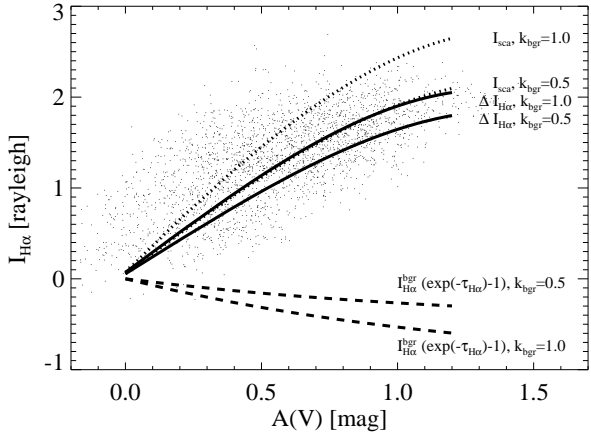}}
\caption{ Intensity of H$\alpha$ surface brightness as a function of
          visual extinction. Dots are the observed values of intensity
          difference $\Delta I_{H\alpha}$.  Dotted line: $I_{sca}$.
          Dashed line: $I_{H\alpha}^{bgr}(\exp(-\tau_{H\alpha})-1)$.
          Solid line: sum of the previous two lines, i.e., $\Delta
          I_{H\alpha}$. }
\label{fig6}
\end{figure}

\subsection{ Comparison of H$\alpha$, $I(100\,\mu$m), $A_V$, and
             644/677\,nm maps } 
Inspection of Fig.~\ref{fig1} shows that the oval cloud at the center
of the images is visible in extinction (panel a), in scattered light
caused by dust (panel b), and in thermal emission by dust (panel
c). The same cloud is also seen in the light of the H$\alpha$ line
(panel d).  The maximum excess H$\alpha$ surface brightness of the
cloud above the background is about 2.0 rayleighs (R), which is
similar to the brightness of the other H$\alpha$ scattering clouds,
1--3\,R, as discovered by Mattila et~al.\ (\cite{mattila07}). We show
in Sect.~3.8 that the detected H$\alpha$ radiation can be explained
solely in terms of scattered radiation.

In addition to the central oval and the SW leg that are seen in the
$A_V$ data, the H$\alpha$ map shows a blob-like structure at
coordinates $l=295\degr$, $b=-26\degr$, which is not seen in any of
the other maps.  The maximum surface brightness of the blob is about
7\,R, and thus the blob cannot be scattered light (see Sect.~3.8).
The emission must instead be {\it in situ} emission from hot, ionized
gas, possibly unconnected to the G294-24 dust cloud.  Similar small
size ($\la 1\degr$) H$\alpha$ enhancements at latitudes
$|b|\ge10\degr$ were reported by Reynolds et~al.\ (\cite{reynolds05})
in the northern sky. Many of them have no obvious embedded or
associated ionizing star. This is also the case for our H$\alpha$
blob.  This H$\alpha$ enhancement could be associated with a planetary
nebula, or be ionized by either a O or early B type star, or a hot
evolved low-mass star (Reynolds et~al.\ \cite{reynolds05}).  The
SIMBAD database does not list any of these kinds of objects within a
distance of 1\degr\, from the center of the blob. The nearest O or
early B type star in SIMBAD is the B2 type star \object{CD-80 228}
($l=292.58\degr$, $b=-27.43\degr$), identified as \object{EC 06387-8045}, 
at a distance of 4.4\,kpc (Kilkenny et~al.\ \cite{kilkenny95}). 
Association of the blob with the star CD-80~228 would mean that the 
blob is located much further away than G294-24 (see Sect.~3.5).

The correlation plots between H$\alpha$, $I(100\,\mu$m),
I(644\,nm/677\,nm), and $A_V$ data are shown in Fig.~\ref{fig2}, after
subtracting the background values from each dataset. To correlate data
sets produced in a compatible way, both sets of data in panel
Fig.~\ref{fig2}{\bf c} are from the original SHASSA survey (Gaustad
et~al.\ \cite{gaustad01}), and analyzed as described above in
Sect.~2.2.  The absolute intensity calibration performed by our
analysis method in Fig.~\ref{fig2}{\bf c} is inaccurate, as indicated
by a calibration difference of a factor of two between the
Finkbeiner's (\cite{finkbeiner03}) (WHAM calibrated) H$\alpha$ data
(panels {\bf a} and {\bf b}) and H$\alpha$ data in panel {\bf c}.
However, the linear relation in panel {\bf c} supports the idea that
the H$\alpha$ intensity is also mainly scattered light.  In the
following, our analysis of the H$\alpha$ surface brightness is based
solely on Finkbeiner's (\cite{finkbeiner03}) data, because its
calibration is based on the accurately calibrated WHAM survey (Haffner
et~al.\ \cite{haffner03}).

\subsection{Temperature and column density of dust}
Equilibrium temperature of dust particles along the ISOSS slew was
derived from the ISOSS and IRIS far-IR data, after subtracting an
estimate of the background intensity, assumed to be a mean value over
the circular region shown in Fig.\ref{fig1}c. We assume a frequency
dependence of $\nu^{2}$ for the emissivity index.  The ISOSS data were
convolved to the resolution of the IRIS data, which is given by a
Gaussian FWHM=4\arcmin.  The temperature along the ISOSS slew (see
Fig.~\ref{fig1}c) is plotted in Fig.~\ref{fig3} as a solid line,
showing that the temperature decreases from about 18.7\,K at the
northern edge of the cloud to a minimum of about 17.5\,K at the
center.

The 100\,$\mu$m optical depth, derived with the formula
\begin{equation}
  \tau(100\,\mu m)=I_{\nu}(100\,\mu m)/B_{\nu}(T_{\mathrm{dust}})
\end{equation}
is shown in Fig.\ref{fig3} as a dotted line. The maximum optical depth
is $\tau$(100\,$\mu$m)$\approx$2\,10$^{-3}$. The distribution of $A_V$
along the same slew is shown in Fig.~\ref{fig3} as a dash-dotted
line. The maximum of $\tau$(100\,$\mu$m) is located to the south of
the 100\,$\mu$m surface brightness maximum, at a position where $A_V$
has its maximum.

To determine the dust temperature from the DIRBE maps, we first
subtracted the estimated background intensities from the DIRBE maps,
derived in the circular area shown in Fig.~\ref{fig1}.  The dust
temperature was then derived by fitting the 100\,$\mu$m, 140\,$\mu$m,
and 240\,$\mu$m intensities with a modified black body having a
$\nu^2$ emissivity law. The minimum temperature of the cloud is
$\sim16.6\pm1.5$\,K, in agreement with the minimum temperature derived
from ISOSS and IRAS data.  It is obvious that we are unable to resolve
any possible colder condensations in the cloud due to the low
resolution of DIRBE data.  Figure~\ref{fig4} shows a map of the
100\,$\mu$m optical depth based on DIRBE data, derived using Eq.1.

The total mass (gas plus dust) of the cloud was calculated with
the formula
\begin{equation}
M_{\mathrm{fir}} = \frac{ I({\mathrm 100\,\mu m})\, D^{2}\,
                         m_{\mathrm H}\, \mu } 
                       { B(100\,\mu \mathrm{m}, T_{\mathrm d})\,
                         \sigma^{\mathrm H}(100\,\mu \mathrm{m}) },
\end{equation}
where $I_{\mathrm 100\,\mu m}$ is the observed flux density, $D$ is
the distance, $m_{\mathrm H}$ is the mass of a hydrogen atom, $\mu$ is
the mean molecular weight (1.4) with respect to the mass of hydrogen
atom, accounting for 10\% helium by number,
$B$(100\,$\mu$m,$T_{\mathrm d}$) is the black-body emission at
temperature $T_{\mathrm d}$, and $\sigma^{\mathrm H}$ is the
absorption cross section per H-nucleus for which we used the value
2.4$\times 10^{-25}$\,cm$^2$ (Lehtinen et~al.\ \cite{lehtinen07}, the
case L1642~C in Table~1).  The total mass of the cloud is about
1000\,M$_{\sun}$ for the distance of 100\,pc.

\subsection{Visual extinction}
The visual extinction map is shown in Fig.~\ref{fig1}a.  Maximum
visual extinction is $\sim$1.2\,mag over the background.  The typical
error in the extinction map is 0.2\,mag.

We can use the visual extinction to estimate the total hydrogen column
density, $N(H) \equiv N(HI)+2N(H_2)$.  As a starting point, we adopt
the value $N(H)/E(B-V)=5.8\,10^{21}$\,cm$^{-2}$\,mag$^{-1}$ for
diffuse clouds (Bohlin et~al.\ \cite{bohlin78}), together with
$A_V/E(B-V)=3.1$ (``diffuse dust'') to obtain
$N(H)/A_V\equiv\beta=1.9\,10^{21}$\,cm$^{-2}$\,mag$^{-1}$

The total cloud mass can then be derived with the formula
\begin{equation}
M = D^2\, \mu\, \beta\, m_{\mathrm H}\, \sum A_V\, dA, 
\end{equation}
where $D$ is the cloud distance, $m_{\mathrm H}$ is the hydrogen mass,
$dA$ is the area of one pixel in the $A_V$ map in steradian, and the
summation is evaluated over the whole cloud (inside the ellipse in
Fig.~\ref{fig1}{\bf a}).  This procedure gives a total mass of
$\sim550$\,M$_{\sun}$.

\subsection{$A_V$ versus $\tau$(100\,$\mu$m)}
Figure~\ref{fig7} shows the relation between visual extinction and
100\,$\mu$m optical depth, after the $A_V$ map has been convolved to
the resolution of the DIRBE data. The slope of the fit gives the
emissivity $\epsilon$=1.5$\pm$0.3\,10$^{-3}$\,mag$^{-1}$. For a
$\nu^{-2}$ emissivity law, the emissivity at 200\,$\mu$m is
4.0$\pm$0.8\,10$^{-4}$\,mag$^{-1}$.  The observed value agrees with
theoretical values of $\epsilon(200\mu$m) for diffuse interstellar
matter (D\'esert et~al.\ \cite{desert90}; Dwek et~al.\ \cite{dwek97b};
Cambr\'esy et~al.\ \cite{cambresy01}; Li \& Draine \cite{li01}; see
Lehtinen et~al.\ \cite{lehtinen07} for a compilation of values of
$\epsilon(200\mu$m)).

\subsection{Distance based on Hipparcos data}
Figure~\ref{fig5} shows $E(B-V)$ versus distance for the stars shown
in Fig.~\ref{fig1}{\bf a}. Color excess $E(B-V)$ has been converted
into $A_V$ by assuming a normal reddening law, $A_V/E(B-V)=3.1$.  For
stars that are within the cloud area and have $E(B-V)>0.1$, the
minimum distance is $\sim100$\,pc. Thus, we adopt a distance of
100\,pc for G294-24, which is less than the distance of 150\,pc to the
adjacent Chamaeleon region (Knude \& H{\O}g \cite{knude98}).

\subsection{HI emission line data}
Figure~\ref{fig8} shows a HI spectrum from the LAB survey towards the
position $l=294.0\degr$, $b=-24.5\degr$. Velocity-longitude diagrams
show that the strength of the narrow component at
$\sim-9$\,km\,s$^{-1}$ follows the intensity of the far-IR emission of
G294-24.  The component at $\sim4$\,km\,s$^{-1}$ is from gas in the
Galactic plane. We fit the spectrum in Fig.~\ref{fig8} with three
Gaussian functions, and use the fit as a template for fits at other
Galactic coordinates; the width of the narrow line at
$\sim-9$\,km\,s$^{-1}$ is kept constant, and the velocities of the
narrow components at $\sim-9$\,km\,s$^{-1}$ and $\sim4$\,km\,s$^{-1}$
are allowed to vary by $\pm3$\,km\,s$^{-1}$.  Figure~\ref{fig9} shows
a map of the line area of the narrow component at
$\sim-9$\,km\,s$^{-1}$. The oval G294-24 cloud and the SW leg are
seen. The blob, seen at coordinates $l=295.0\degr$, $b=-26\degr$ on
the $H\alpha$ map, cannot be seen as a separate entity in the LAB
data, lending support to the idea that the blob is ionized gas.  The
maximum of line area is located on the eastern side of the cloud, in
contrast to the maximum of 100\,$\mu$m optical depth, which is located
on the western side.

The column density (atoms per cm$^{-2}$) of atomic hydrogen can be
obtained from (see e.g., Verschuur \cite{verschuur74})
\begin{equation}
N(HI) = 1.82\,10^{18} \int T_B\, dv, 
\end{equation}
where $T_B$ is the brightness temperature and the integral is line
area in units of K\,km\,s$^{-1}$, and low optical depth is assumed.
The total atomic hydrogen mass of the cloud, integrated over the
ellipse in Fig.~\ref{fig9}, is about 41\,M$_{\sun}$. Thus, the atomic
hydrogen mass of the cloud is 4\%-7\% of the total mass of the cloud.

The large-scale $^{12}$CO($J=1-0$) survey of the Chamaeleon region by
Mizuno et~al.\ (\cite{mizuno01}) shows a weak, isolated region near
the center of G294-24, seen only in the velocity range
2--6\,km\,s$^{-1}$.  The integrated intensity of $^{12}$CO($J=1-0$)
emission towards the cloud is estimated to be $\sim2$\,K\,km\,s$^{-1}$
(Figure.~1 of Mizuno et~al.). The possible relation of the CO emission
with the cloud G294-24 has to be studied with observations of higher
angular resolution.

\subsection{IRAS or 2MASS point sources}
We checked the IRAS point source catalog for objects that have
spectral energy distribution typical of young stellar objects.  There
are several 100\,$\mu$m only sources in the cloud. They are probably
small cirrus structures seen as points sources by IRAS at 100\,$\mu$m,
and we do not consider them further here. The only source that has
fluxes measured at least at the two longest IRAS wavelengths is
\object{IRAS 08048-8211}, which was identified as a galaxy by Buta
(\cite{buta95}).

We compiled a color-color diagram ($J-H$ versus \ $H-K$ magnitudes)
for all the stars with magnitude errors smaller than 0.05\,mag at $J$,
$H$, and $K$ band. There is no star within the cloud area that
exhibits a significant infrared excess above the color indices that
can be explained by interstellar reddening.

Based on the non-existence of IRAS point sources and 2MASS objects
with colors characteristic of young stellar objects, we conclude that
the cloud is devoid of star formation.

\subsection{ Radiative transfer calculations }
The maximum possible surface brightness of any Galactic dust cloud,
due to scattering, is limited by the average all-sky H$\alpha$ surface
brightness of $\sim$8\,R.  Since the H$\alpha$ excess surface
brightness of G294-24 of $\sim2.4$\,R is well below this value, we
conclude that it can be explained solely by scattered radiation.  To
verify this assumption, we simulated scattered H$\alpha$ radiation
with Monte Carlo radiative transfer calculations (Mattila
\cite{mattila70}; Juvela \& Padoan \cite{juvela03}; Juvela
\cite{juvela05}).  We used the WHAM Northern Sky Survey (Haffner
\cite{haffner03}) to derive the intensity of the northern H$\alpha$
background sky illuminating the model cloud.  The missing southern sky
was recreated by assuming symmetry about the Galactic latitude and
longitude. The physical model of the cloud is a spherical, homogeneous
cloud with $A_V=1.2$\,mag of visual extinction through the cloud
center. Properties of dust particles are based on Draine's
(\cite{draine03}) ``Milky Way'' dust model, with albedo $a=0.67$ and
asymmetry parameter $g=0.5$ at the wavelength of the H$\alpha$ line.
For more details of applying Monte Carlo method to the scattering of
H$\alpha$ radiation, we refer to Mattila et~al.\ (\cite{mattila07}).

A general formula for the differential surface brightness of H$\alpha$
light towards a dust cloud, measured over the brightness of the
adjacent sky, is
\begin{equation}
\Delta I_{H\alpha} = I_{H\alpha}^{bgr}\,\exp(-\tau_{H\alpha}) + I_{sca} + 
                I_{H\alpha}^{fgr} - (I_{H\alpha}^{bgr} + I_{H\alpha}^{fgr}),
\end{equation}
where the first term on the right-hand side is the intensity of
emission coming from behind the cloud attenuated by the optical depth
through the cloud, the second term is the intensity of radiation
scattered off the cloud, the third term is the intensity of emission
between the cloud and the observer, and the term in parenthesis is the
intensity of sky adjacent to the cloud. The radiative transfer
calculations provide us with the value of $I_{sca}$ as a function of
optical depth through the cloud.

We have no a priori information about the relative values of
$I_{H\alpha}^{bgr}$ and $I_{H\alpha}^{fgr}$. Thus, in the
model we use a free parameter, $k_{bgr}$, which is defined as 
\begin{equation}
k_{bgr} \equiv \frac{ I_{H\alpha}^{bgr} } 
                  { I_{H\alpha}^{bgr} + I_{H\alpha}^{fgr} }, \; 0<k_{bgr}<1.
\end{equation}
Figure~\ref{fig6} shows the observed intensity difference, $\Delta
I_{H\alpha}$, and the results of radiative transfer calculations for
two values of $k_{bgr}$.  The dotted lines show the scattered
intensity $I_{sca}$, the dashed lines show the calculated
$I_{H\alpha}^{bgr}(\exp(-\tau_{H\alpha})-1)$, and the thick solid line
is the sum of them both, i.e., $\Delta I_{H\alpha}$.  The dashed lines
indicate the surface brightness of the cloud, relative to the
background sky, in the absence of scattering ($a=0$); the cloud would
be seen in absorption.  The case $k_{bgr}=1.0$ provides a good fit to
the data. Because of the high noise level of the data, we cannot
determine an accurate value for the parameter $k_{bgr}$, although a
value of about 0.5 must be a lower limit.  Thus, we verify that the
observed H$\alpha$ surface brightness can be explained solely by
scattered radiation. The agreement also shows that the adopted dust
scattering parameters of the Draine (\cite{draine03}) model are
realistic.

In addition, we used the above-mentioned model cloud in radiative
transfer calculations of continuum radiation (Juvela \& Padoan
\cite{juvela03}), giving us the surface brightness of the cloud at
far-infrared wavelengths. The interstellar radiation field surrounding
the cloud is taken from Mathis et~al.\ (\cite{mathis83}). Properties
of dust particles are based on Draine's (\cite{draine03}) ``Milky
Way'' dust model, not including stochastically heated dust grains.
The derived 100\,$\mu$m, 170\,$\mu$m, and 240\,$\mu$m maximum surface
brightnesses are about 12\,MJy\,sr$^{-1}$, 24\,MJy\,sr$^{-1}$, and
19\,MJy\,sr$^{-1}$, respectively.  The observed maximum surface
brightnesses above the background are about 13\,MJy\,sr$^{-1}$,
25\,MJy\,sr$^{-1}$, and 21\,MJy\,sr$^{-1}$, respectively.  We then
derived the dust temperature using the 100\,$\mu$m and 170\,$\mu$m
maps. The minimum temperature is about 17.6\,K, in good agreement with
the temperature derived from IRIS and ISOSS data, of about 17.5\,K
(see Sect.~3.2).

\section{Conclusions}
We have studied the general properties of an undocumented large,
nearby (distance $\sim100$\,pc) translucent cloud. The cloud is
without star formation, and thus illuminated only by the general
interstellar radiation field.  We have measured a drop in the
temperature of the dust particles of about 1.2\,K between the edge and
the center of the cloud, the minimum dust temperature being
$\sim$17.5\,K.  The total mass of the cloud is about
550--1000\,M$_{\sun}$. The cloud is detected in the 21\,cm HI
spin-flip line, and the atomic hydrogen mass of the cloud is about
40\,M$_{\sun}$.  The excess of diffuse H$\alpha$ surface brightness of
the cloud over the background sky can be naturally explained as the
general interstellar radiation field being scattered off the dust
grains.

\begin{acknowledgements}
The work of K.L., M.J.\ and K.M.\ has been supported by the Finnish
Academy through grants Nos.\ 1204415, 1210518 and 1201269, which is
gratefully acknowledged.  This publication makes use of data products
from the Two Micron All Sky Survey, which is a joint project of the
University of Massachusetts and the Infrared Processing and Analysis
Center/California Institute of Technology, funded by the National
Aeronautics and Space Administration and the National Science
Foundation.  This publication uses data from the Southern H-Alpha Sky
Survey Atlas (SHASSA), which is supported by the National Science
Foundation.  The Wisconsin H-Alpha Mapper is funded by the National
Science Foundation.  This research has made use of SAOImage DS9,
developed by Smithsonian Astrophysical Observatory
\end{acknowledgements}

\end{document}